\documentclass[aps,prb,twocolumn,showpacs,groupedaddress]{revtex4}  % for review and submission
\usepackage{graphicx}  % needed for figures
\usepackage{dcolumn}   % needed for some tables
\usepackage{bm}       % for math
\usepackage{amssymb}   % for math
\usepackage{amsmath}
\usepackage{mathtools}
\usepackage{color}

\begin{document}

% the following line is for submission, including submission to the arXiv!!
%\hspace{5.2in} \mbox{Fermilab-Pub-04/xxx-E}

\title{Pressure-induced Lifshitz transition in NbP: Raman, x-ray diffraction, electrical transport and density functional theory}
\author{Satyendra Nath Gupta$^1$, Anjali Singh$^{2 \dagger}$, Koushik Pal$^{2 \dagger}$,  D.V.S.Muthu$^1$, C. Shekhar$^3$, Yanpeng Qi$^3$, Pavel G. Naumov$^3$, Sergey A. Medvedev$^3$, C. Felser$^3$, U. V. Waghmare$^2$ and A. K. Sood$^1$}
\thanks{To whom correspondence should be addressed}
\email{asood@physics.iisc.ernet.in}
\affiliation{Department of Physics, Indian Institute of Science, Bangalore-560012,India $^2$Theoretical Sciences Unit, Jawaharlal Nehru 
Centre for Advanced Scientific Research, Bangalore-560064,India  $^3$Max Planck Institute for Chemical Physics of Solids, 01187 Dresden, 
Germany $^\dagger$These two authors contributed equally}

\date{\today}

\begin{abstract}
We report high pressure Raman, synchrotron x-ray diffraction and electrical transport studies on Weyl semimetals NbP and 
TaP along with first-principles density functional theoretical (DFT) analysis. The frequencies of 
first-order Raman modes of NbP harden with increasing pressure and exhibit a slope change at P$_c$ $\sim$ 9 GPa, and its resistivity exhibits a minimum at P$_c$. The pressure-dependent volume of NbP exhibits a  change in its bulk modulus from 207 GPa to 243 GPa at P$_c$. Using DFT calculations, we show that these anomalies are associated with pressure induced Lifshitz transition which involves appearance of electron and hole pockets in its electronic structure. In 
contrast, results of Raman and synchrotron x-ray diffraction experiments 
on TaP and DFT calculations show that TaP is quite robust under pressure and does not undergo any phase transition. 

\end{abstract}
%\pacs
\maketitle
\section{Introduction}
Soon after the discovery of Dirac equation in 1928, it was pointed out that another massless solution of the Dirac equation represents a new 
kind of particle called Weyl Fermions\cite{weyl1929elektron}. The recent developments in topological insulators and topological semimetals 
have opened a way to realize Weyl Fermions in terms of low energy excitations where two non-degenerate linear dispersion electron bands 
cross each other at isolated points (called Weyl nodes) in Brillouin zone\cite{PhysRevX.5.011029}. This band structure has been observed in Weyl 
semimetals (WSMs)\cite{xu2015discovery,lv2015experimental,lv2015observation,yang2015weyl,xu2015experimental}. The Weyl nodes in WSMs always come in 
spatially separated pairs of opposite chirality, making WSMs different from Dirac semimetals which have two degenerate Weyl nodes that form one 
Dirac node due to time-reversal and inversion symmetry. WSMs exhibit many exotic properties induced by Weyl nodes like topological surface states 
with Fermi arcs \cite{lv2015experimental,xu2015discovery,yang2015weyl} and a negative longitudinal magnetoresistance due to the chiral 
anomaly\cite{huang2015observation, zhang2016signatures,shekhar2015extremely,nielsen1983adler,aji2012adler,son2013chiral,kim2013dirac,hosur2013recent}. 
Recently, the non-centrosymmetric TaAs, TaP, NbAs, and NbP have been predicted to be the candidate materials for WSMs with twelve pairs of Weyl nodes 
in their 3D Brillouin zones\cite{PhysRevX.5.011029,huang2015weyl}. The experimental evidence of Weyl nodes has been observed in this family by topological 
surface state and bulk electronic band structure measurements using angle resolved photoemission spectroscopy 
(ARPES)\cite{liu2016evolution,xu2015discovery,Nature Physics 2015}.

The Fermi surface topology of Weyl semimetals can be modified by a small change in the Fermi energy, due to their low intrinsic charge 
carrier densities. Application of pressure is known to be a powerful approach to tune the electronic and lattice structure of the material. High pressure 
synchrotron x-ray study of TaAs up to 53 GPa along with ab initio calculations\cite{zhou2016pressure} show that TaAs goes to hexagonal P$\bar{6}$m2 phase 
at 14 GPa from the ambient I4$_1$md phase, along with changes in the electronic states. Further high pressure magneto-transport study up to 
2.3 GPa of NbAs\cite{luo2016hard} shows that the Fermi surface exhibits an anisotropic evolution under pressure. Similarly magneto-transport study 
of NbP up to 2.8 GPa\cite{dos2016pressure} shows significant effect on the amplitudes of Shubnikov–de Haas oscillations due to the subtle changes in the 
Fermi surface. The effect of pressure on lattice and phonons has not been studied in NbP as well as in TaP. The objective of the present study is to 
examine NbP and TaP under pressure (carried out upto 25 GPa) using Raman spectroscopy and synchrotron x-ray diffraction as well as electrical 
transport (for NbP). DFT calculations have been done to gain microscopic insight into pressure effect. Our main results for NbP are: (i) The phonon 
frequencies of first order Raman modes exhibit slope change at P$_c$ $\sim$ 9 GPa, (ii) The resistivity of NbP pass through a 
minimum at P$_c$, (iii) The bulk modulus shows an increase  at  P$_c$, (iv) The DFT calculations reveal that these anomalies at P$_c$ are due to 
Lifshitz transition involving  appearance of electron and hole pockets in its electronic structure. In comparison to NbP, our results on TaP show 
no phase transition up to 27 GPa.

\section{Experimental details}
High-quality single crystals of NbP and TaP were grown via a chemical vapor transport reaction using iodine as a transport agent\cite{Martin 1988}. 
Initially, polycrystalline powder of NbP was synthesized by a direct reaction of niobium (Chempur 99.9\%) and red phosphorus 
(Heraeus 99.999\%) kept in an evacuated fused silica tube for 48 hours at 800 $^o$C. Starting from this microcrystalline powder, the single-crystals of NbP were synthesized by chemical vapor transport in a temperature gradient starting from 850  $^o$C (source) to 950 $^o$C (sink) and 
a transport agent with a concentration of 13,5 mg/cm$^3$ iodine (Alfa Aesar 99,998\%). Similar route was used for the growth of TaP crystals. Raman experiments were carried out at room temperature on these 
well characterized single crystals of NbP and TaP \cite{shekhar2015extremely,Arnold 2016} using confocal Horiba 800 spectrometer, peltier cooled CCD and 532 nm 
diode laser. The high pressure synchrotron x-ray diffraction measurements were done at DESY on the beam line (P02) using 0.2888\AA ~x-ray radiation. 
Pressure was generated using a diamond anvil cell (DAC), where a thin platelet ($\sim 100 \mu m$) of single crystals of NbP/TaP, was embedded in a 
4:1 methanol: ethanol  pressure transmitting medium along with a ruby chip for pressure calibration into a stainless steel gasket inserted between 
the diamonds. The electrical resistivity at different pressures was measured by the direct current van der Pauw technique in an in-house-designed 
diamond-anvil cell equipped with diamond anvils with a 500 $\mu$m culet. A cleaved NbP single crystal of a suitable size was cut and placed into 
the central hole of a tungsten gasket with an insulating cubic BN/epoxy layer without pressure transmitting medium. The electrical leads 
fabricated from 5 $\mu$m thick Pt foil were attached to the sample. The pressure was again determined by the ruby luminescence method.

\section{Computational Details}

Our first-principles calculations are based on density functional theory (DFT) as
implemented in Quantum ESPRESSO package \cite{qe}, in which the interaction between ionic
cores and valence electrons is modeled with ultrasoft pseudopotentials \cite{USPP}.
The exchange-correlation energy of electrons is treated within a Local Density Approximation (LDA) using a functional
form parameterized by Perdew-Zunger \cite{PZ}. We use an energy cutoff of 60 Ry to truncate the
plane wave basis for representing Kohn-Sham wave functions, and energy cutoff of 600 Ry on the basis set
to represent charge density. We relaxed structures to minimize energy till the magnitude of Hellman-Feynman
force on each atom is less than 0.03 eV/\AA. In self-consistent Kohn-Sham (KS) calculations  with 
primitive cells, the Brillouin zone (BZ) integrations were sampled with a
uniform mesh of 11x11x11 k-points. 
We determined electronic structure by including the spin-orbit coupling (SOC) through
use of relativistic ultrasoft pseudopotentials\cite{Corso}.
We determined dynamical matrices and phonons at wavevectors on a 2x2x2 mesh in the BZ using DFT linear response 
(Quantum ESPRESSO implementation based on Green's function).
We used scalar relativistic pseudopotential
for phonon calculations as inclusion of SOC changed phonon frequencies by  a very small value ($<$ 1 cm$^{-1}$) 
(Fig.~\ref{fig2}c, d). 
To determine the bulk modulus (B), we have fitted the P vs V plot to Birch Murnaghan equation of state.
\begin{align}
P(V)=\frac{3B}{2}[({\frac{V_{0}}{V}})^{\frac{7}{3}}-({\frac{V_{0}}{V}})^{\frac{5}{3}}] [1+\frac{3}{4}(B^{'}-4)(({\frac{V_{0}}{V}})^{\frac{2}{3}}-1)]
\end{align}

where V is the volume at  pressure P, V$_0$ is the reference volume,  B is the bulk modulus, and $B^{'}$ is the pressure derivative of the bulk modulus.
Fermi surfaces are interpolated on dense k-grids (equivalent to 33x33x33 in the whole BZ).

\section{Results and discussions}

\subsection{Experimental Results}
\begin{figure}
    \begin{center}
       \includegraphics[width=0.5\textwidth]{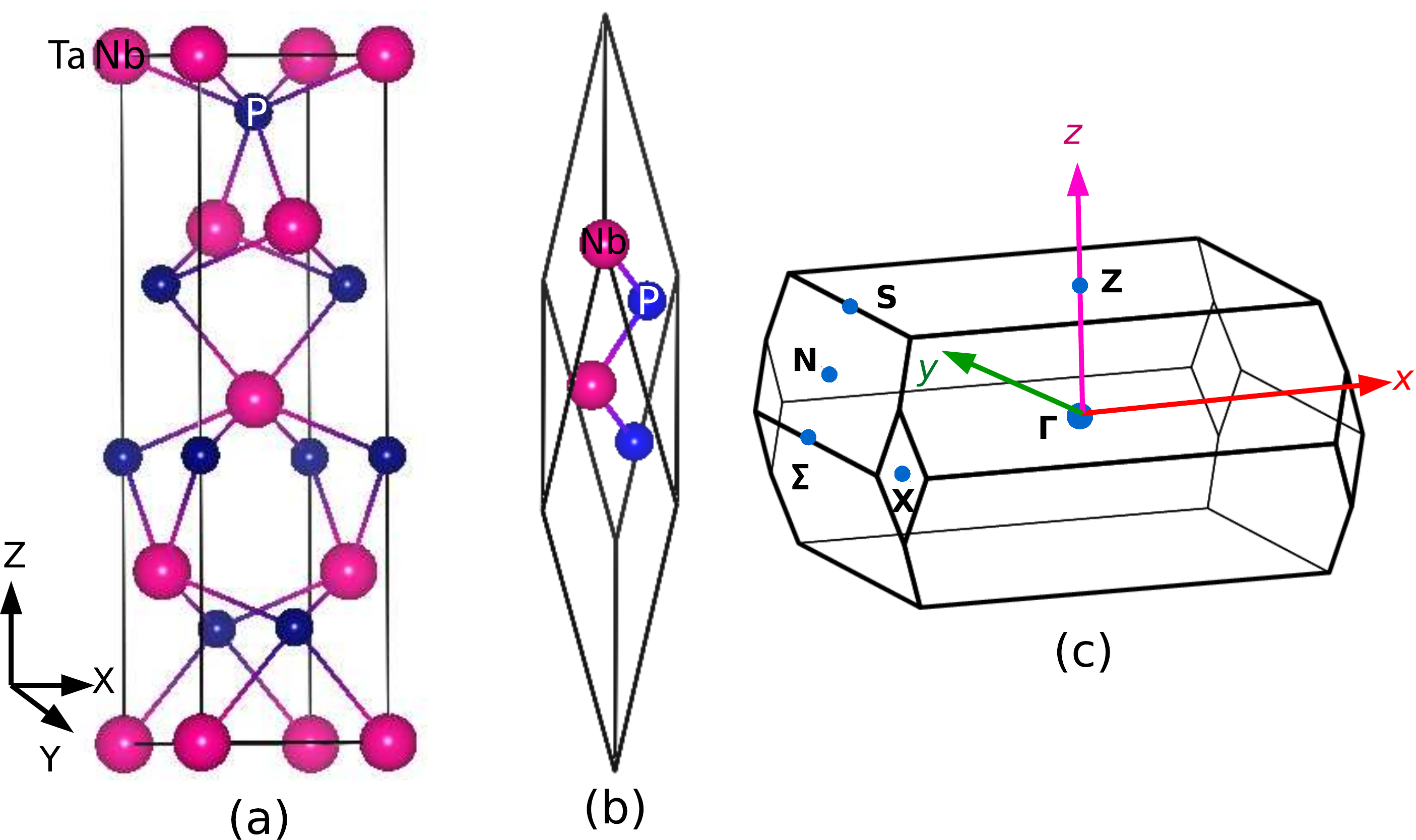}
       \caption{(Color online) Crystal structure of bulk NbP and TaP with (a) conventional, (b) primitive unit cells and (c) their Brillouin zone (BZ).}
   \label{structure}
   \end{center}
\end{figure}

NbP and TaP have the same geometric structure (non-centrosymmetric body centered tetragonal (BCT) structure), which belongs to the 
space group of \textit{I${4_1}$md} ($C^{11}_{4v}$, no. 109)\cite{Boller} with two atoms each of transition metal and phosphorous 
per primitive cell (see Fig.~\ref{structure}a, b). In this structure, Nb/Ta and P atoms have six nearest neighbors in a trigonal prismatic
coordination. To understand the crystal structure, we can consider a stack of square lattice planes, where each plane is formed of either
Nb/Ta or P atoms. Each plane is shifted by half a lattice constant in either direction (x or y) relative to the one below it. Due to these 
shifts, a screw pattern arises in the lattice along the z-direction leading to a
non-symmorphic $C_4$ rotational symmetry that requires a translation along the z direction by c/4. It is clear from this crystal structure 
that NbP and TaP lack the spatial inversion symmetry.
If time-reversal symmetry is respected, the broken inversion symmetry is an important condition for the Weyl semimetallic
phase.

\begin{figure}[t]
\includegraphics[width=0.5\textwidth]{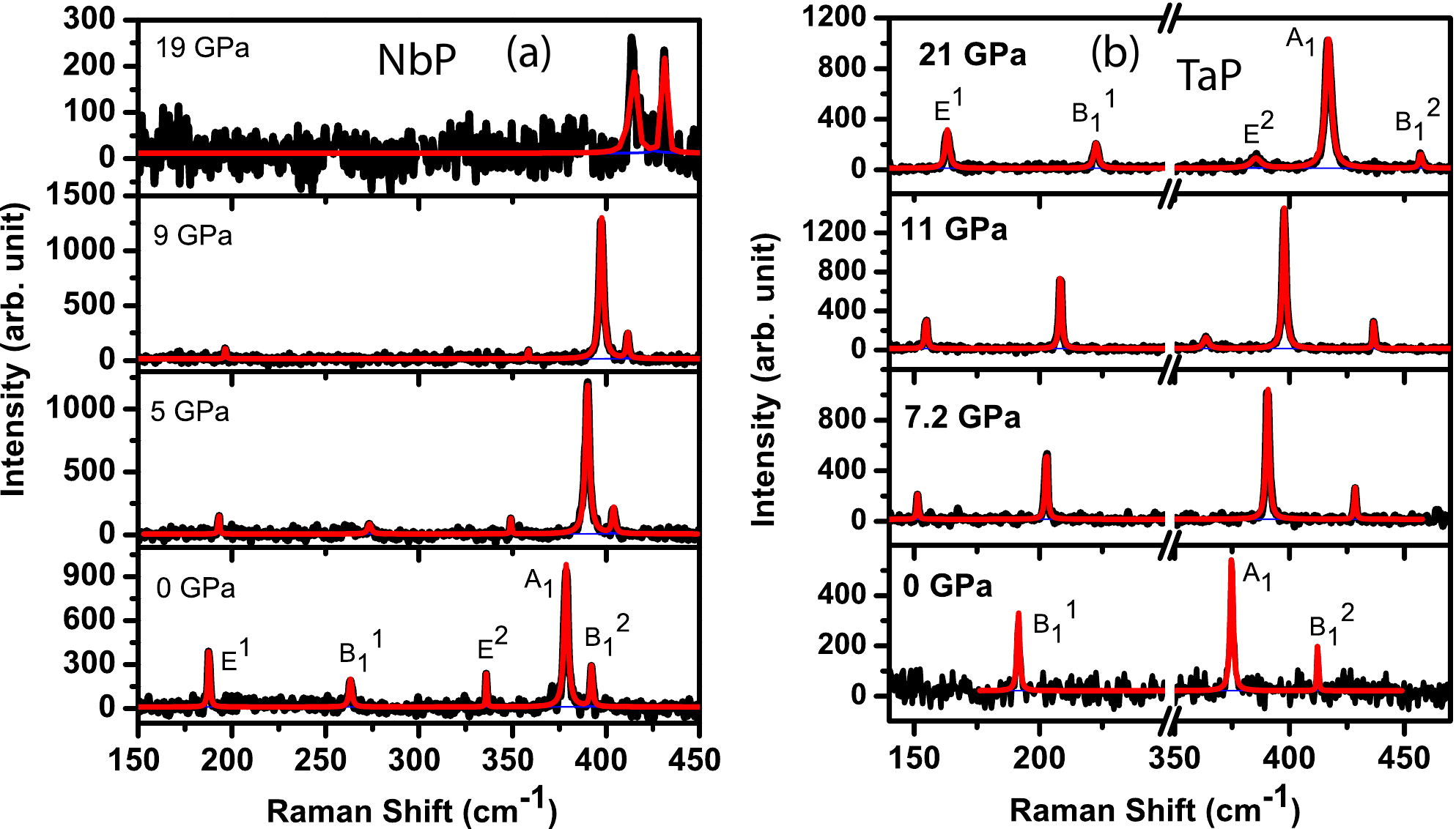}
\caption{(Color online) Raman spectrum of NbP (a) and TaP (b) at a few representative pressures. The solid lines (red and blue) are the Lorentzian fit to the measured specta (black). 
\label{Fig-RamanNbP}}
\end{figure}

\begin{figure*}[t]
\includegraphics[width=0.9\textwidth]{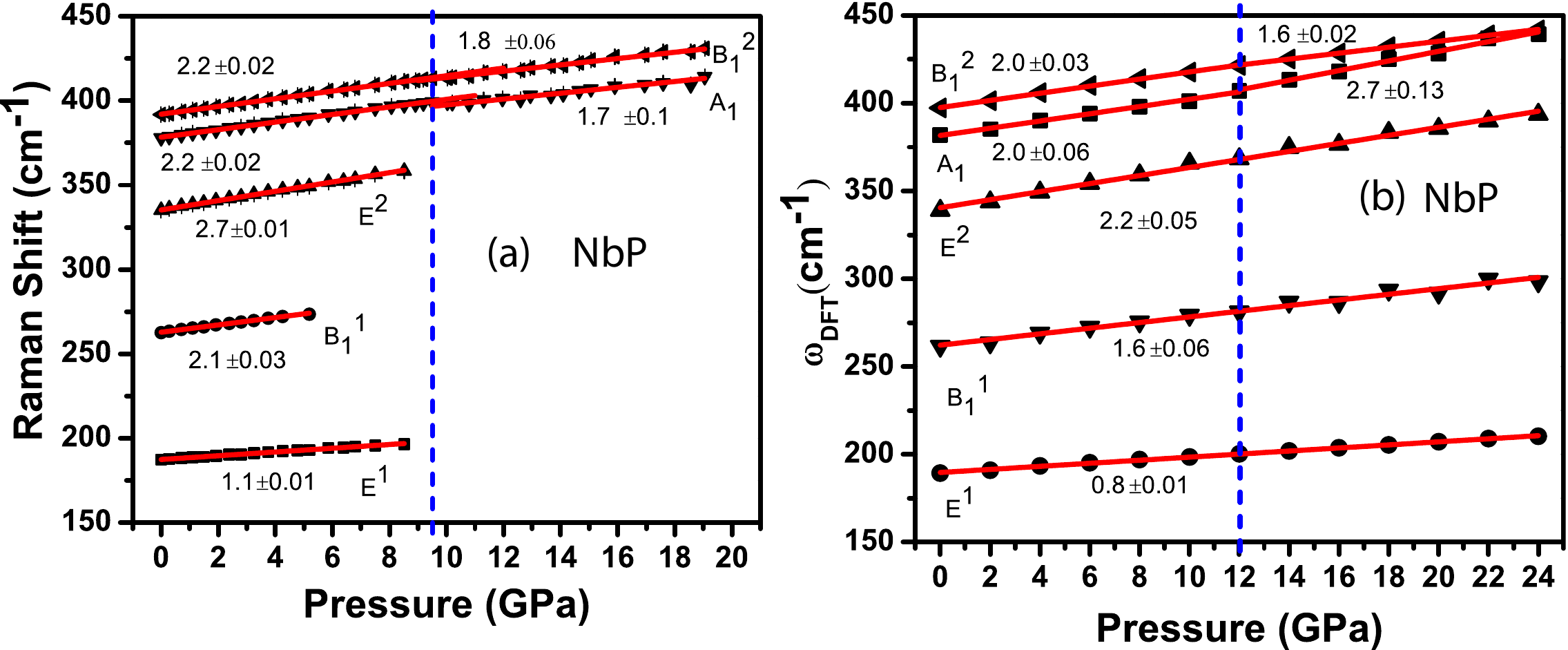}
\caption{(Color online) Pressure dependence of phonons of NbP: (a) Experiment; (b) Theory. The solid lines are 
the linear fit to the data. The slope S in the unit of cm$^{-1}$/GPa is given near each of the 
the lines. The vertical dashed lines mark the transition pressure P$_c$}.
\label{Fig-phononNbP}
\end{figure*}

\begin{figure}[t]
\includegraphics[width=0.45\textwidth]{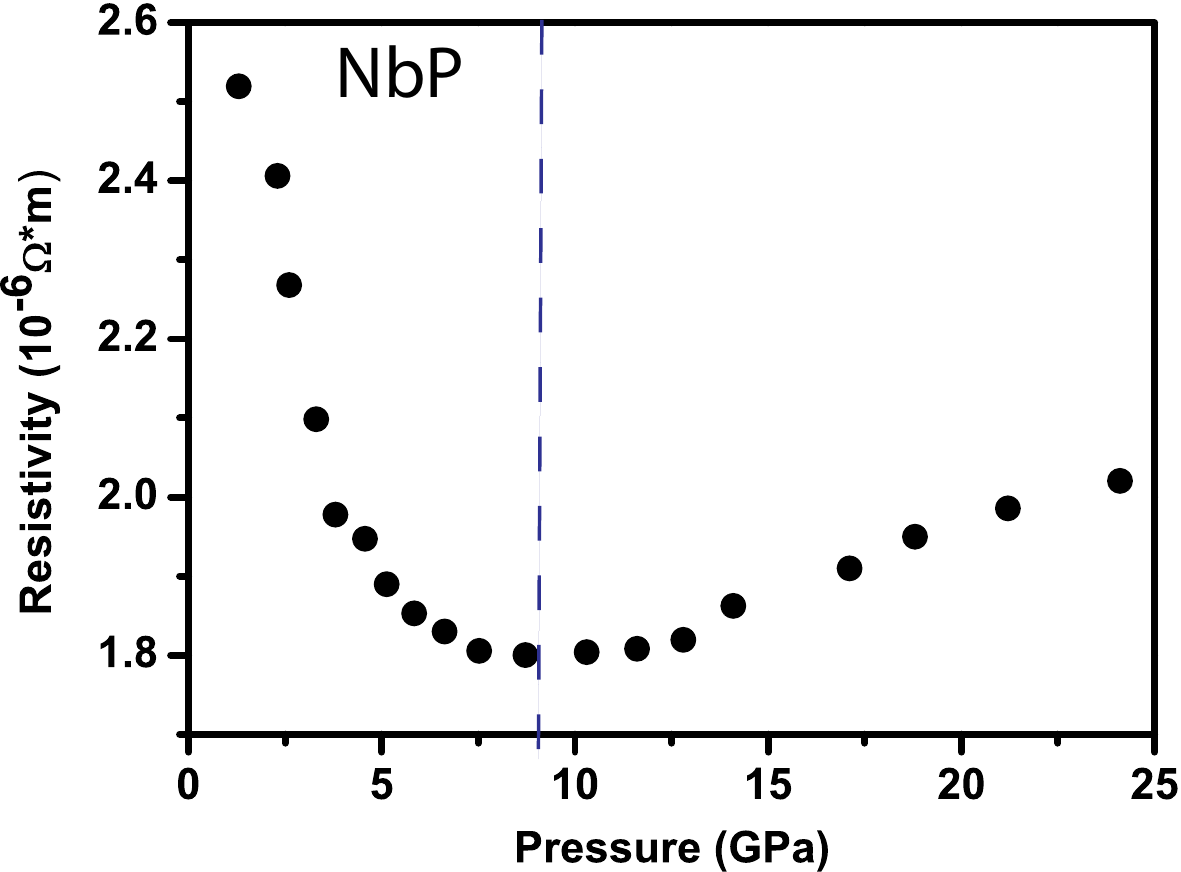}
\caption{(Color online) Measured resistivity of NbP as a function of pressure. The resistivity passes through a minimum at $\sim$ 9GPa as indicated by a vertical dashed line.
\label{Fig-resistanceNbP}}
\end{figure}

\begin{figure*}[t]
\includegraphics[width=0.8\textwidth]{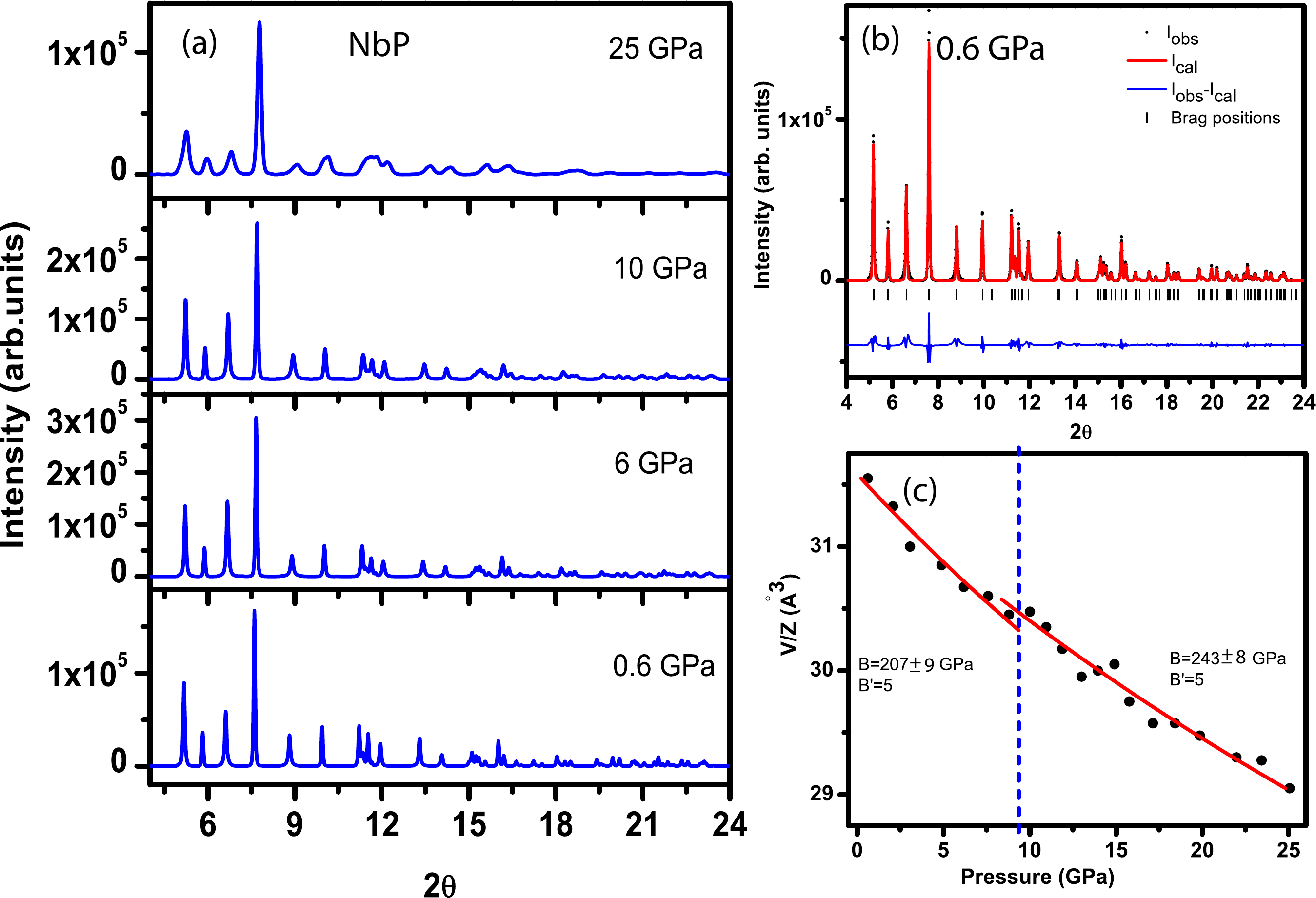}
\caption{(Color online) (a) X-ray diffraction pattern of NbP as a function of pressure. (b) The Rietveld refinement 
of X-ray diffraction pattern at 0.6 GPa pressure. (c) Pressure dependence of volume per formula unit (z) of NbP. The vertical dashed 
line indicates the transition pressure, at which bulk modulus changes. 
\label{Fig-xrdNbP}}
\end{figure*}

According to factor group analysis, this system has [A$_1$+E] acoustic phonons and [A$_1$+2B$_1$+3E] optical phonons. All the optical phonons are 
Raman active\cite{liu2016comparative}. Fig.\ref{Fig-RamanNbP} (a) shows Raman spectra of NbP at a few typical pressures. Five Raman modes are observed 
in NbP and assigned to E$^1$, B$_1^1$, E$^2$, A$_1$ and B$_1^2$ irreducible representations, based on first-principles calculations (discussed latter). 
The Lorentzian line shapes were fitted to the Raman spectra to extract the phonon frequencies and full width at half  maximum (FWHM). The Raman modes 
E$^1$, B$_1^1$ and E$^2$ become very weak after $\sim$ 9 GPa, making it difficult to follow them. The extracted phonon frequencies at different pressures,   plotted in Fig.\ref{Fig-phononNbP} (a), show clearly that the phonon frequencies of A$_1$ and B$_1^2$  modes exhibit  a  change in the slope S ($\dfrac{d\omega}{dP}$) at P$_{c}$ $\sim$ 9 GPa, signifying a phase transition.  To further investigate this transition, we did resistivity  measurements as function of pressure.  Fig.\ref{Fig-resistanceNbP}  shows that the resistivity first decreases with increasing pressure and then increases, showing a minimum at  P$_{c}$.  Thus our Raman and resistivity results as a function of pressure show a phase transition at  P$_{c}$. To find out structural changes, if any, at  P$_{c}$, high pressure x-ray diffraction measurements were carried out.  Fig.\ref{Fig-xrdNbP} (a) shows  x-ray diffraction patterns  at a few representative pressures, revealing  no new diffraction peaks or vanishing of the existing peaks till 25 GPa. This  rules out the possibility of structural change at P$_{c}$. In order to get the lattice parameters, we did Rietveld refinement using Jana 2006 \cite{Petricek} as shown for example at 0.6 GPa in Fig.\ref{Fig-xrdNbP} (b). The diffraction pattern is well fitted by Jana 2006 at all pressures. Fig.\ref{Fig-xrdNbP} (c) shows   pressure dependence of the unit cell  volume per formula unit (z).  The solid red lines are fit to Eq.(1).
It is clear from Fig.\ref{Fig-xrdNbP} (c) that P-V data fitted using Eq. (1) (keeping $B^{'}$  fixed at 5) shows a clear change in bulk modulus: 
B=207 GPa (P $\le$ P$_c$) and 243 GPa (P $\ge$ P$_c$). Thus high pressure x-ray data rules out a structural phase transition at P$_{c}$ and 
suggests an isostructural electronic phase transition\cite{Xiaoa 2010,Zhao 2015,Hong 2016}.

\begin{figure*}[t]
\includegraphics[width=0.9\textwidth]{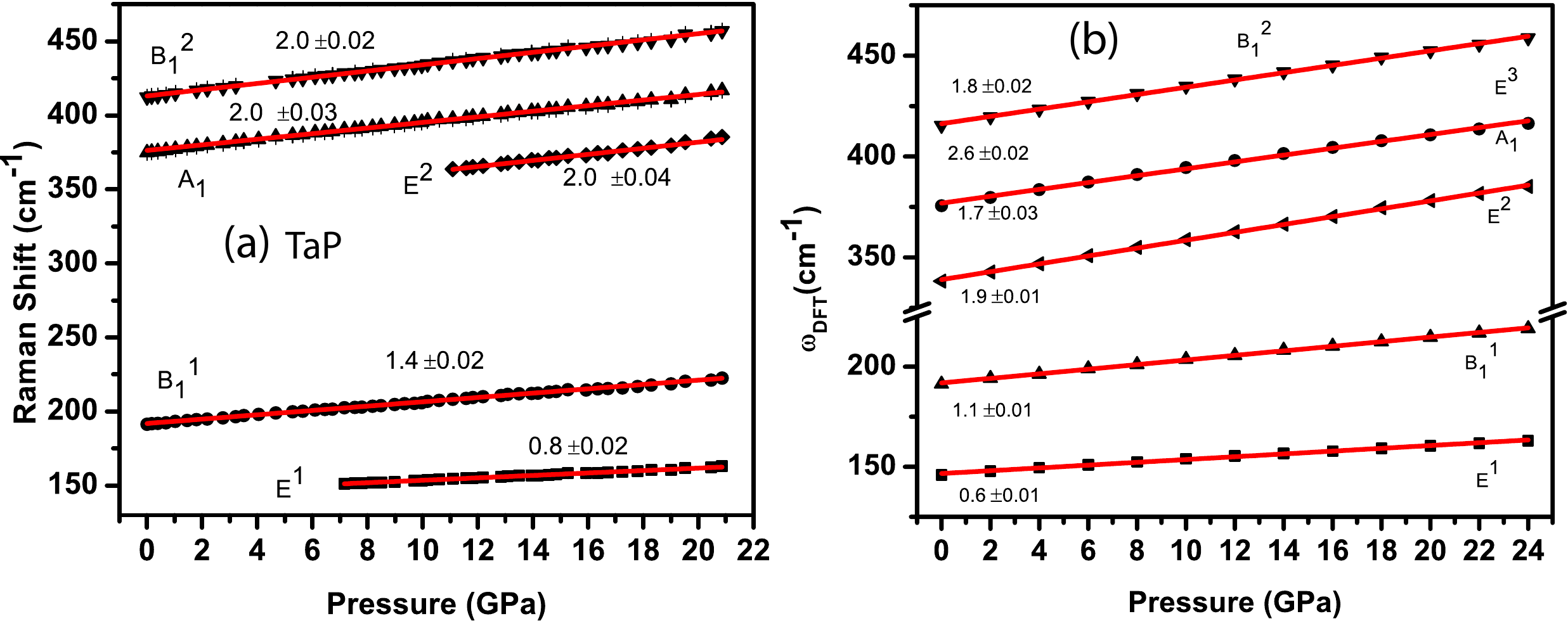}
\caption{(Color online) Pressure dependence of phonons of TaP: (a) Experiments; (b) Theory. The solid lines are the linear fit 
to the experimental data. The slope S in the unit of cm$^{-1}$/GPa is given near the lines.  
\label{Fig-RamanTaP}}
\end{figure*}

\begin{figure*}[t]
\includegraphics[width=0.8\textwidth]{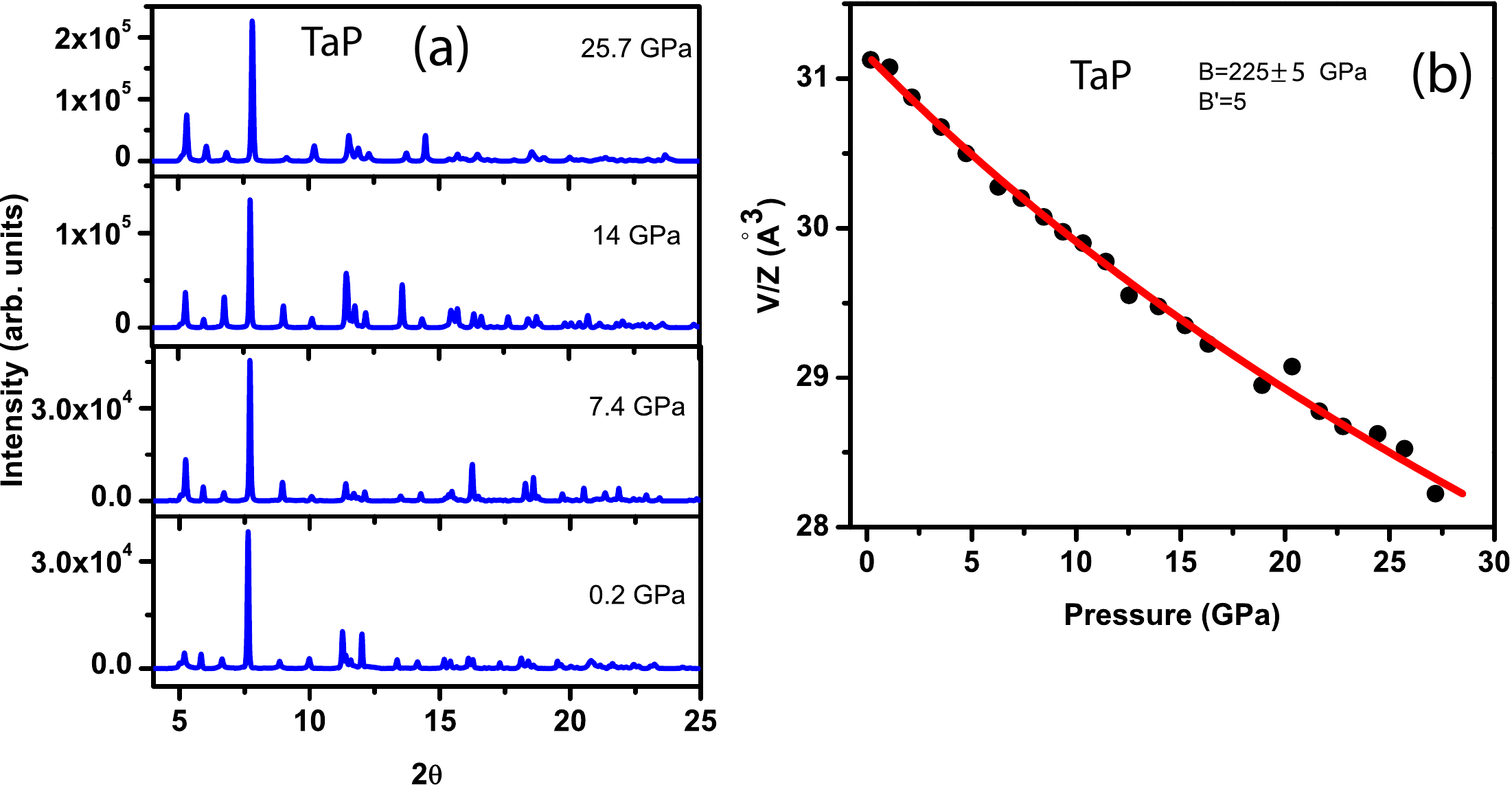}
\caption{(Color online) (a) X-ray diffraction pattern of TaP as function of pressure, and (b) pressure dependence of volume per formula unit (z) of TaP. 
\label{Fig-xrdTaP}}
\end{figure*}

We will now present our results on TaP. Fig.\ref{Fig-RamanNbP} (b) shows  Raman spectra of TaP at a few representative pressures. As before, 
the solid red lines are the Lorentzian function fitted to the experimental data (black lines). At low pressure, only 3 modes could be recorded 
and the two other modes (E$^1$ and E$^2$) could be seen at higher pressures. The zero pressure extrapolated values of the frequencies of 
E$^1$ and E$^2$ modes are close to the values obtained from DFT calculations. The pressure dependence of  phonon frequencies of 
E$^1$, B$_1^1$, E$^2$, A$_1$ and B$_1^2$ modes are shown in Fig.\ref{Fig-RamanTaP} (a). It is clear from Fig.\ref{Fig-RamanTaP} (a) that all 
the Raman modes harden with increasing pressure and there is no slope change with pressure which is in  contrast to the results obtained in NbP. 
Fig.\ref{Fig-xrdTaP} (a) shows the pressure dependence of synchrotron x-ray diffraction patterns of TaP at a few pressures. We note that the number 
of diffraction peaks remain the same  in the entire pressure range, suggesting no structural phase transition. The lattice parameters were obtained 
by Rietveld refinement using Jana 2006\cite{Petricek}. The pressure dependence of the volume per formula unit (z) is shown 
in Fig.\ref{Fig-xrdTaP} (b). The solid red line is the fit to Birch Murnaghan equation of state (Eq.1), showing that there is no change in 
the bulk modulus of TaP over the entire pressure range (Fig.\ref{Fig-xrdTaP} (b)). We can thus conclude that TaP does not show any phase transition 
till 27 GPa.

\subsection{Theoretical calculations}

\begin{figure*}
    \begin{center}
       \includegraphics[width=0.7\textwidth]{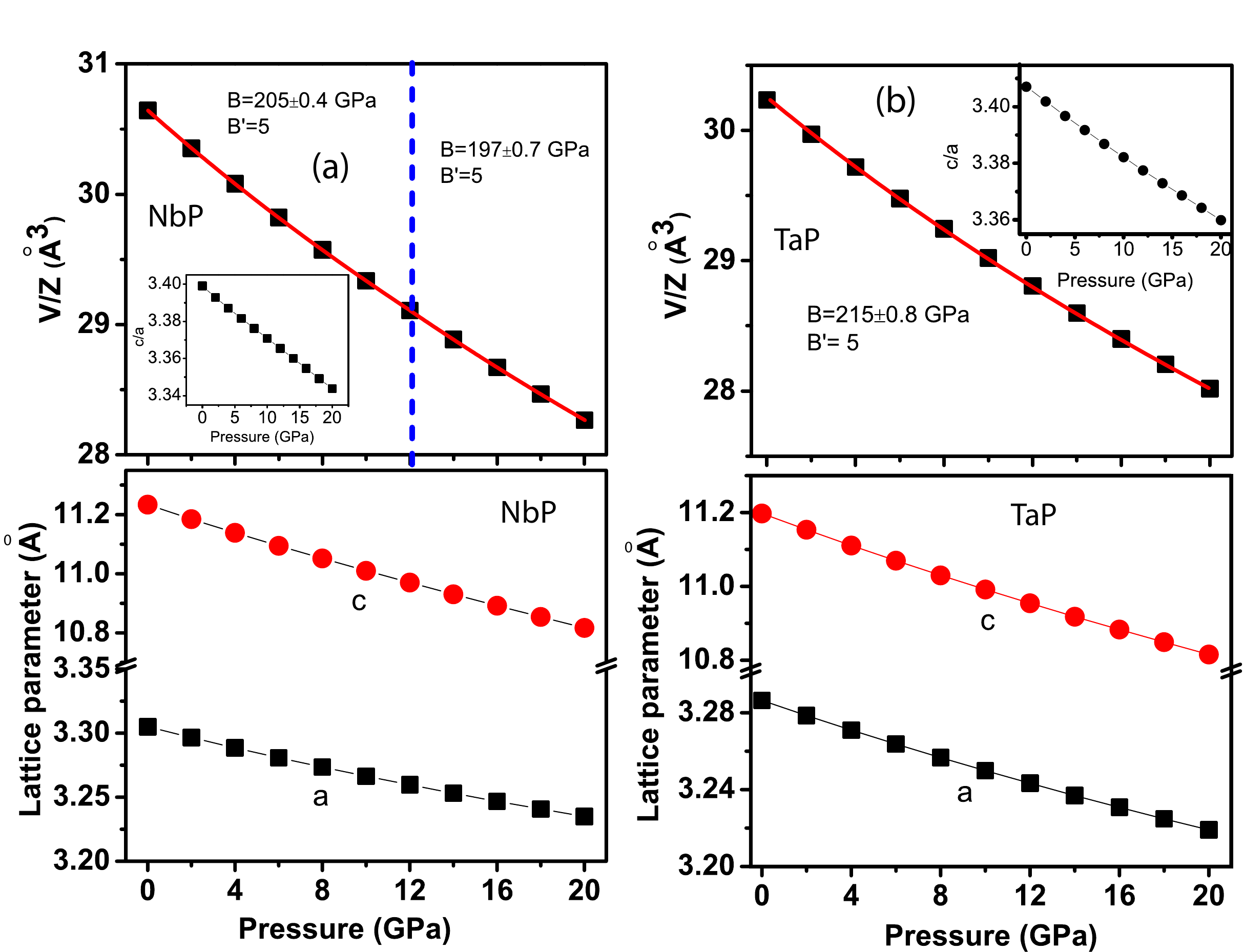}
       \caption{(Color online) Calculated lattice structural parameters a, c, c/a and volume per formula unit of bulk NbP (a) and TaP (b) as a function of pressure.}
   \label{fig1}
   \end{center}
\end{figure*}

The calculated structural parameters are as follows: For NbP: a=b=3.31\AA, c=11.23\AA; for TaP: a=b=3.29\AA, c= 11.2\AA. These values are in good 
agreement with experiments \cite{Lee}. Lattice structural parameters of bulk NbP and TaP show smooth variation as a function of hydrostatic 
pressure (Fig.~\ref{fig1} a,b), signifying the robustness of the crystal structure of NbP and TaP upto 20 GPa. Further, the absence of anomalies 
in c/a ratio  rules out the possibility of an isostructural phase transition. Calculated bulk moduli (from V vs P plot Fig.~\ref{fig1} a,b) of 
the BCT structure of NbP  (TaP) are 205$\pm$0.4 GPa  (215$\pm$0.8 GPa), which agree well with experimental values of 207$\pm$9 GPa (225$\pm$5 GPa). 
Also, there is a  change  in the bulk modulus of NbP at 12 GPa, consistent with the experiment. 

\begin{figure*}
\begin{center}             
\includegraphics[width=0.8\textwidth]{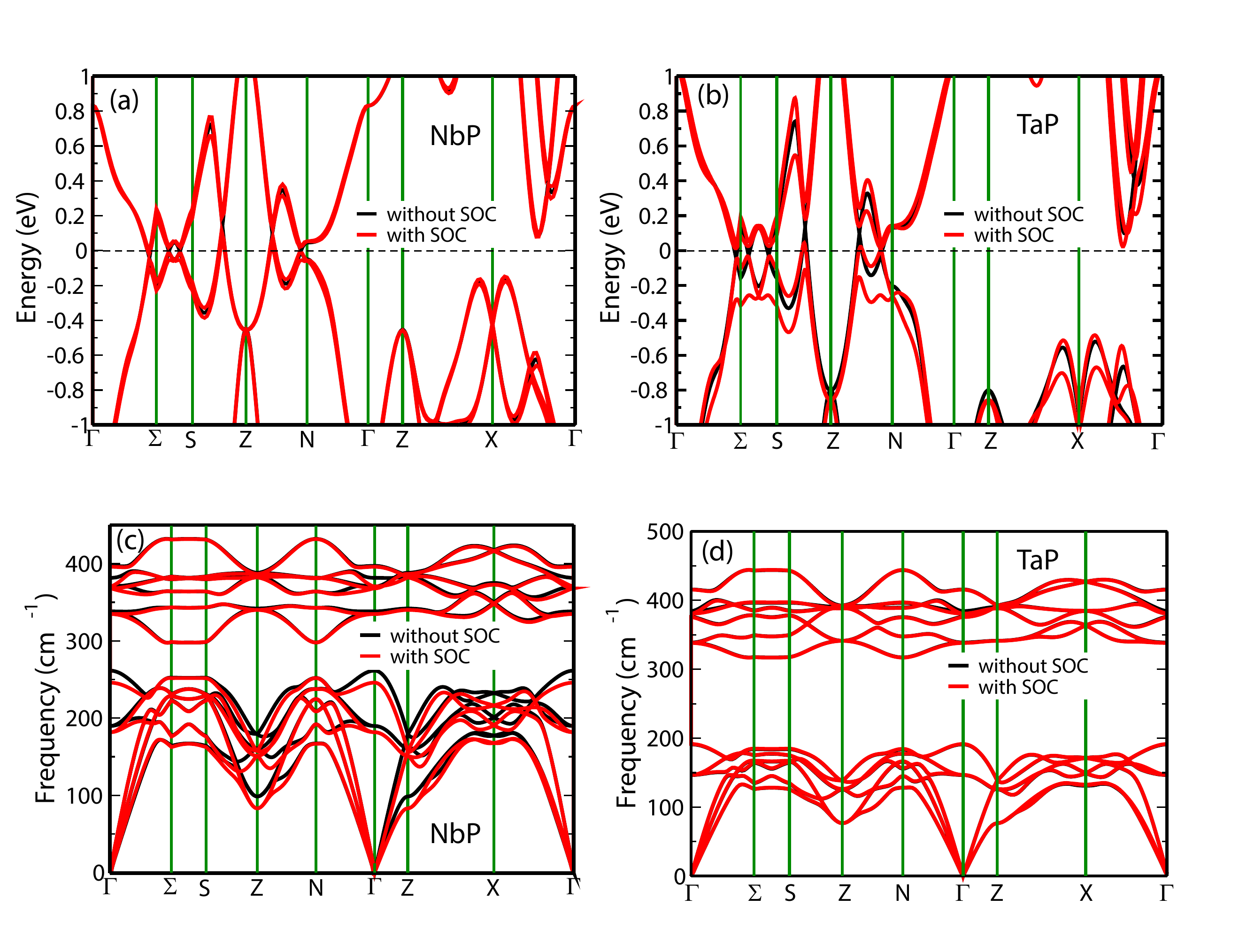} 
\caption{(Color online) Electronic structure (a, b) and phonon dispersion (c, d) of bulk NbP and TaP respectively at P = 0 GPa, calculated with and without inclusion of spin-orbit coupling.}
\label{fig2}
\end{center}
\end{figure*}

The electronic structure and phonon dispersion of NbP and TaP with and without spin-orbit coupling at 0 GPa are shown in Fig.\ref{fig2}. In the absence 
of spin-orbit coupling, the valance band maximum (VBM) and conduction band minimum (CBM) cross the Fermi level along high symmetric paths 
($\Gamma$-$\Sigma$-S-Z-N) of the Brillouin zone (Fig.~\ref{fig2}a, b). This crossing leads to formation of closed ring structure (nodal rings 
\cite{PhysRevX.5.011029}) (in $k_x$ $=$ 0 and $k_y$ $=$ 0 planes) which is based on the fact that bands have opposite eigenvalues of mirror symmetry. As 
spin-orbit coupling is introduced, the nodal rings disappear, \textit{i.e.} gap opens up along these high-symmetry lines 
(refer blue line in Fig.~\ref{fig2}a, b) and Weyl nodes are generated \cite{Lee}. This gap is smaller in NbP  as compared to TaP, as expected 
from a weaker spin-orbit coupling in Nb as compared to Ta compounds. Our calculated  phonon frequencies at zone center are in good agreement 
with experiments (see Table~\ref{tab2}). 

\begin{table}
\centering
\caption{Comparison of the observed and calculated frequencies of Raman active modes of NbP and TaP at 0 GPa. Excellent 
agreement between theory and experiment  is evident.}
\begin{tabular}[t]{c|cc|cc}
\hline

&\multicolumn{2}{c|}{NbP (0 GPa, $\omega$ in cm$^{-1}$)}&\multicolumn{2}{c}{TaP (0 GPa, $\omega$ in cm$^{-1}$)}\\
\hline
Modes &Experiment & Theory & Experiment & Theory \\
\hline
$E^1$& 189 &182&145&146 \\
$B_1^1$& 263& 262&191&192 \\
$E^2$& 335&339&341&339 \\
$B_1^2$& 391&397&411&415 \\
$A_1$& 377& 382&373&375 \\
\hline
\end{tabular}
\label{tab2}
\end{table}

\begin{figure*}        
\includegraphics[width=0.7\textwidth]{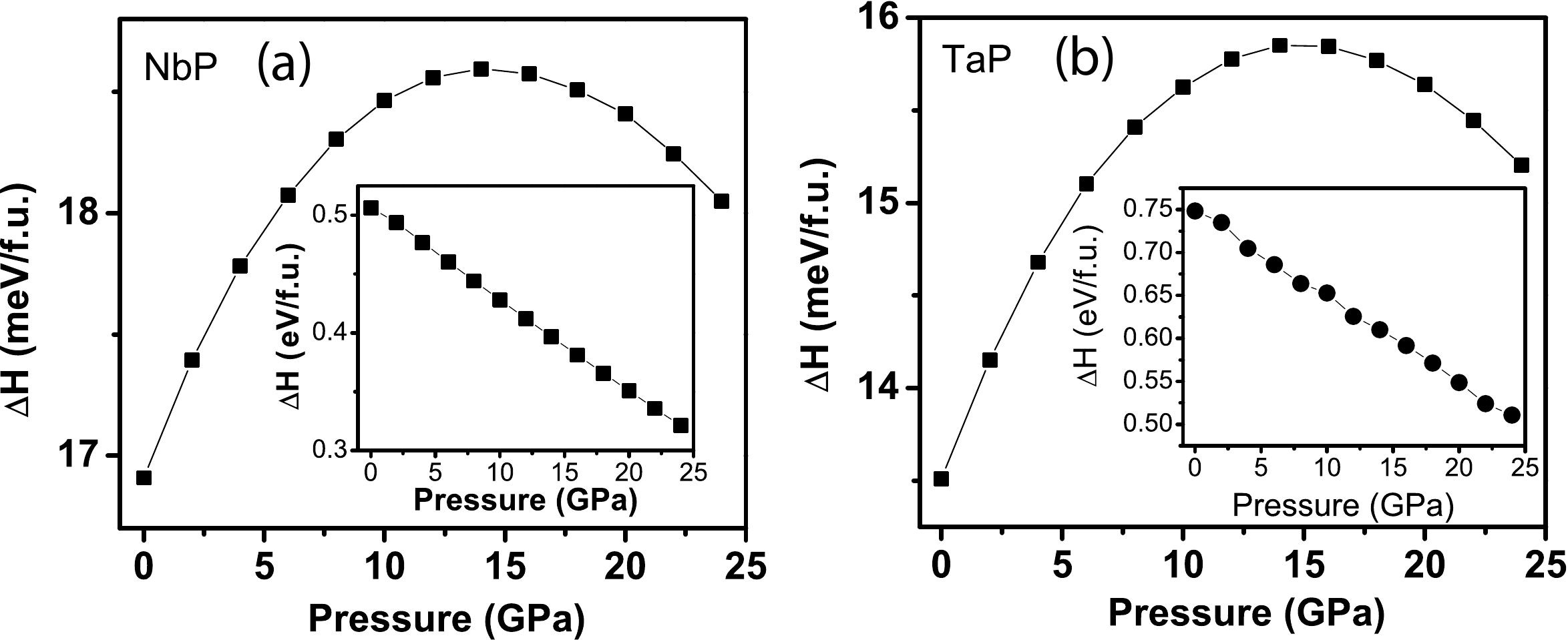}
 \caption{(Color online) Pressure dependent difference in enthalpy of body-centered tetragonal (BCT) and hexagonal (hex) structures of NbP (a) and TaP (b). $\Delta{H}$ of BCT and cubic structures are given in the inset.}  
\label{fig5}
\end{figure*}

Our calculations of phonon frequencies of NbP  as a function of hydrostatic pressure (Fig.~\ref{Fig-phononNbP} (b)) reveal that $A_1$ and $B_1^2$ modes 
exhibit  change in S at P$_c^{theory}$ $\sim$ 12 GPa while the other three modes E$^1$, E$^2$ and $B_1^1$ do not show any slope change till 24 GPa, 
consistent with the experimental results. The difference between the observed transition pressure P$_c$ $\sim$ 9 GPa and the calculated one 
P$_c^{theory}$ $\sim$ 12 GPa is partly due to the errors in calculated equilibrium lattice constants. To identify if it is a the structural phase 
transition, we considered two high symmetry crystal structures: (a) hexagonal (Tungsten Carbide structure with space group P-6m2, No. 162) and (b) 
cubic (CsCl structure with space group Pm-3m, No. 221), and calculated the enthalpy differences between these phases and the low pressure BCT structure. 
Although enthalpy difference ($\Delta$H) between the hexagonal and BCT phase first increases slightly with pressure and then decreases, it does not 
attain negative value (Fig.~\ref{fig5}a). Thus, a structural phase transition from BCT to hexagonal phase is ruled out. Similarly, $\Delta$H, between 
the cubic and BCT phase decreases with pressure but does not crossover to negative value up to 24 GPa (inset of Fig.~\ref{fig5}a), 
ruling out a structural phase transition to cubic phase as well. Our conclusion of the absence of a structural phase transition in NbP with 
pressure is consistent with our x-ray diffraction experiments.

\begin{figure*}         
\includegraphics[width=0.7\textwidth]{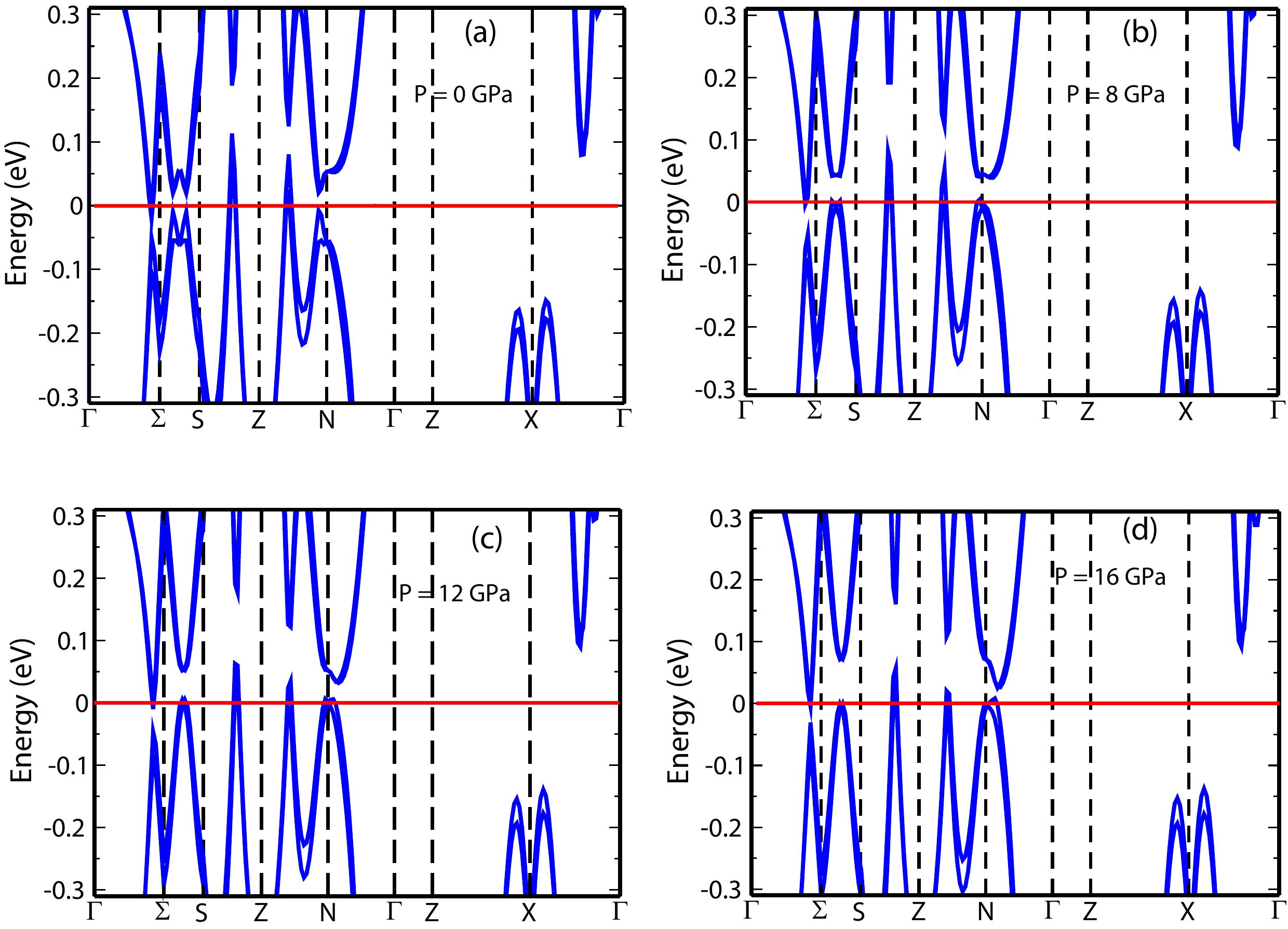}         
 \caption{(Color online) Electronic structure of bulk NbP calculated at (a) 0 GPa, (b) 8 GPa, (c) 12 GPa and (d) 16 GPa.}
\label{fig6}
\end{figure*}

\begin{figure*}   
       \includegraphics[width=0.7\textwidth]{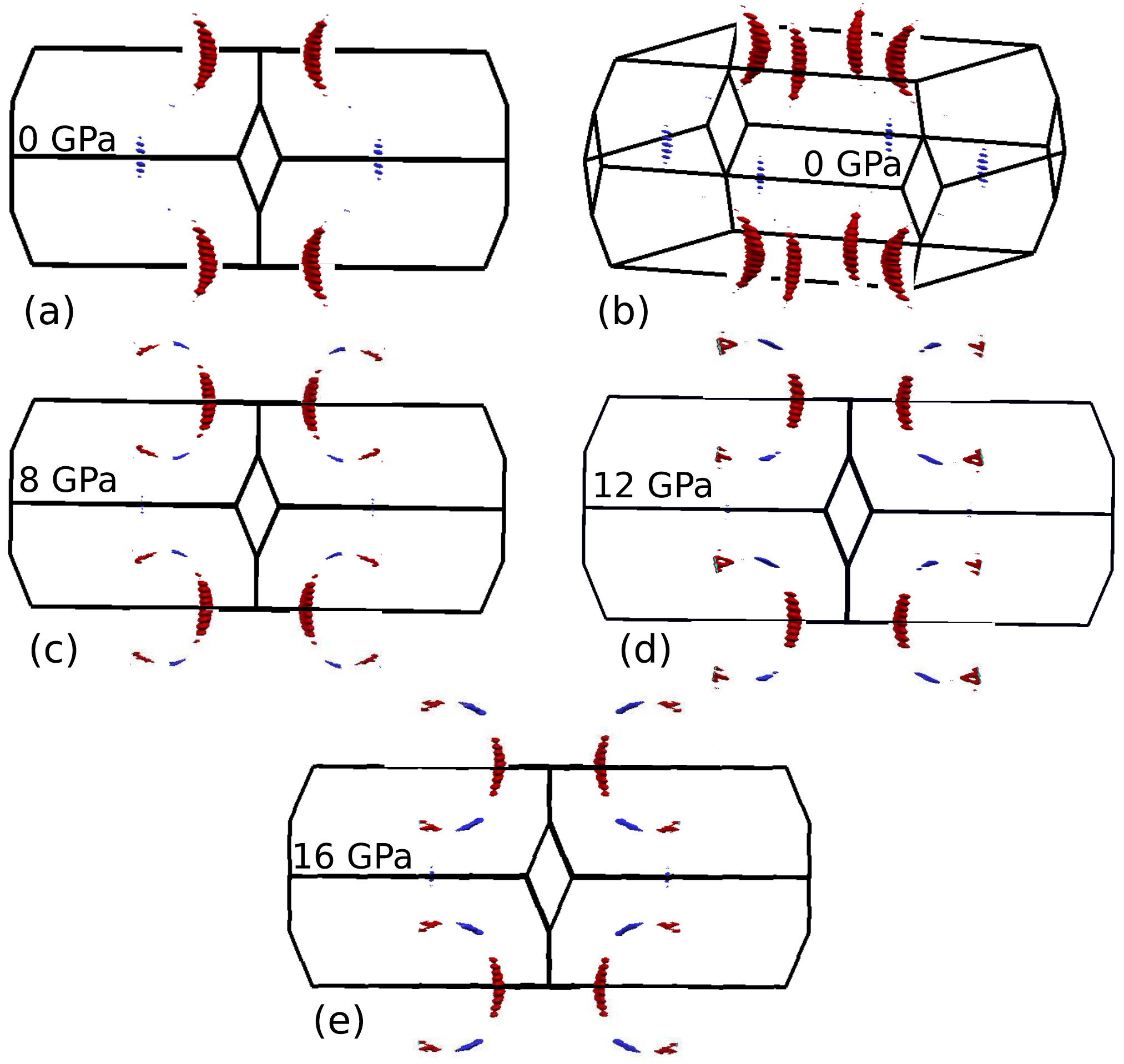}
       \caption{(Color online) Evolution of the Fermi surface of NbP with pressure. Two views of the Fermi surface at 0 GPa (a, b). Fermi surface, at 8 GPa (c), 12 GPa (d) and 16 GPa (e) show changes in size of electron and hole pockets with pressure. Red color
shows hole pockets whereas blue color shows electron pockets.}
   \label{fig7}  
 \end{figure*}

To understand the origin of change in pressure coefficients of Raman active modes ($A_1$ and $B_1^2$ modes) of NbP above P$_c^{theory}$=12 GPa, we study the evolution of  electronic structure with pressure. At P = 0 GPa, NbP is semimetallic in nature 
\textit{i.e } finite density of states at Fermi level. Near the Fermi energy, electronic structure of NbP at 0 GPa shows 
presence of electron pocket along $\Gamma$-$\Sigma$ and hole pockets along S-Z and Z-N lines
(refer Fig.~\ref{fig6}a). As the hydrostatic pressure does not alter the symmetry of the crystal, energy levels do not split, but those 
near the Fermi energy change notably giving rise to pressure induced transfer of electrons from one pocket to another in order
to maintain the total number of carriers (\textit{i.e.} size of electron and hole pockets changes with pressure) \cite{anjali}. 
Interestingly, at 12 GPa a small hole pocket appears along $\Gamma$-N path (see Fig.~\ref{fig6}c). To probe this further, we monitored 
the evolution of Fermi surface with pressure, particularly the electron (red surface) and hole (blue surface) 
pockets (Fig.~\ref{fig7}a). Electron and hole pockets are almost semicircular and distributed along the rings \cite{PhysRevX.5.011029, huang2015weyl} on the 
$k_x = 0$ and $k_y = 0$ mirror planes in the BZ (Fig.~\ref{fig7}b). At 8 GPa, the size of electron pocket in vicinity of N-point 
reduces and a set of electron-pockets (in between existing hole pocket and one appearing near N-point) and hole-pockets (in vicinity of N-point) 
 appear (see Fig.~\ref{fig7}c). Further increase in pressure results in changes of
the shape and size of 
electron and hole pockets (refer Fig.~\ref{fig7}). Note that the electron pocket in vicinity of $\Sigma$-point 
(Fig.~\ref{fig7}a) does not disappear completely 
with increasing pressure. It is thus clear that there is notable changes in topology of Fermi surface with pressure. 
Since, the Fermi surface changes with applied pressure without breaking the structural symmetry, we assign it as a Lifshitz transition 
 occurring at P$_c^{theory}$. We note that there are changes in S  at 12 GPa (Fig.~\ref{Fig-phononNbP}b). Thus, there is a clear correlation between the pressure coefficients 
of Raman active modes and electronic phase transition.
The small changes in the electronic band topology or Fermi surface topology are driven by external parameters and reflected 
in the anomalies in the measurable quantities of two types: (a) the appearance or disappearance of electron and hole 
pockets and (b) the rupturing of necks connecting Fermi-arcs. In the present work, we observe Lifshitz transition\cite{Lifshitz} associated
with the former.

We now present our results for pressure dependence of Raman active modes of TaP. 
We show the variation of Raman active modes as a function of  pressure in Fig.~\ref{Fig-RamanTaP} (b).
Calculated Raman active modes do not exhibit any change in S,
consistent with our experimental results on TaP (Fig.~\ref{Fig-RamanTaP} (a)).
To further check whether TaP would undergo a structural phase transition to either the hexagonal or cubic
phases, we estimated the enthalpies of BCT, cubic and hexagonal crystal structures
of TaP. 
Although  the enthalpy difference  ($\Delta H$) (inset of Fig.~\ref{fig5}b)  between the cubic and BCT
phases decreases with pressure, it does not become negative even up to 24 GPa. Thus,  a structural
phase transition to the cubic phase of TaP  is ruled out. Similarly, ($\Delta H$) between
the  hexagonal and BCT  phases (Fig.~\ref{fig5}b) also does not exhibit a crossover to negative values, signifying absence of any structural 
phase transition, consistent with ours experiments. 
                                                                                       
\section{Summary}
Frequencies of A$_1$ and B$_1^2$ modes of NbP show a change in slope (S = $\dfrac{d\omega}{dP}$) 
at P$_{c}$ and its resistivity exhibits a minimum at  P$_{c}$. The pressure dependent volume of NbP reveals that 
there is a change in the bulk modulus by $\sim$ 36 GPa at  P$_{c}$. We show that these anomalies 
are associated with pressure-induced Lifshitz transition  at P$_{c}$ using first-principles density 
functional theoretical analysis. The BCT structure of NbP is robust as a function of pressure upto 24 GPa. 
Our experimental measurements and theoretical analysis of TaP show that its 
BCT structure is quite robust under pressure, and exhibits no structural or electronic phase 
transitions  with pressure.  
 
\section{Acknowledgments}
AKS thanks Department of Science and Technology, India for the financial support. AS and KP are thankful to Jawaharlal Nehru Centre 
for Advanced Scientific Research, India for research fellowship. UVW acknowledges support from Indo-Korea Science and Technology (IKST) Center 
and a J. C. Bose National Fellowship of the Department of Science and Technology, Govt of India.

%\bibliography{ref}{} 

\begin{thebibliography}{22}
\expandafter\ifx\csname natexlab\endcsname\relax\def\natexlab#1{#1}\fi
\expandafter\ifx\csname bibnamefont\endcsname\relax
  \def\bibnamefont#1{#1}\fi
\expandafter\ifx\csname bibfnamefont\endcsname\relax
  \def\bibfnamefont#1{#1}\fi
\expandafter\ifx\csname citenamefont\endcsname\relax
  \def\citenamefont#1{#1}\fi
\expandafter\ifx\csname url\endcsname\relax
  \def\url#1{\texttt{#1}}\fi
\expandafter\ifx\csname urlprefix\endcsname\relax\def\urlprefix{URL }\fi
\providecommand{\bibinfo}[2]{#2}
\providecommand{\eprint}[2][]{\url{#2}}

\bibitem[{\citenamefont{Weyl}(1929)}]{weyl1929elektron}
\bibinfo{author}{\bibfnamefont{H.}~\bibnamefont{Weyl}},
  \bibinfo{journal}{Zeitschrift f{\"u}r Physik A Hadrons and Nuclei}
  \textbf{\bibinfo{volume}{56}}, \bibinfo{pages}{330} (\bibinfo{year}{1929}).

\bibitem{PhysRevX.5.011029}
H. Weng, C. Fang, Z. Fang, B. A. Bernevig, and X. Dai, Phys. Rev. X {\bf 5}, 011029 (2015).


\bibitem{xu2015discovery}
S. Y. Xu, I. Belopolski, N. Alidoust, M. Neupane, G. Bian, C. Zhang, R. Sankar, G. Chang, Z. Yuan, C. C. Lee, S. M. Huang, H. Zheng, J. Ma, D. S. Sanchez, B. Wang, A. Bansil, F. Chou, P. P. Shibayev, H. Lin, S. Jia, and M. Z. Hasan, Science {\bf 349}, 613 (2015).
  
  

\bibitem{lv2015experimental}
B. Q. Lv, H. M. Weng, B. B. Fu, X. P. Wang, H. Miao, J. Ma, P. Richard, X. C. Huang, L. X. Zhao, G. F. Chen, Z. Fang, X. Dai, T. Qian, and H. Ding, Phys. Rev. X {\bf 5},  031013 (2015).
  
  
  
  

\bibitem{lv2015observation}
B. Q. Lv,	N. Xu,	H. M. Weng,	J. Z. Ma,	P. Richard,	X. C. Huang,	L. X. Zhao,	G. F. Chen,	C. E. Matt,	F. Bisti,	V. N. Strocov,	J. Mesot,	Z. Fang,	X. Dai,	T. Qian,	M. Shi	and H. Ding, Nat. Phys. {\bf 11}, 724 (2015).




\bibitem{yang2015weyl}
L. X. Yang,	Z. K. Liu,	Y. Sun,	H. Peng,	H. F. Yang,	T. Zhang,	B. Zhou,	Y. Zhang,	Y. F. Guo,	M. Rahn,	D. Prabhakaran,	Z. Hussain,	S.-K. Mo,	C. Felser,	B. Yan	and Y. L. Chen, Nat. Phys. {\bf 11}, 728 (2015).



\bibitem{xu2015experimental}
S. Y. Xu, I. Belopolski, D. S. Sanchez, C. Zhang, G. Chang, C. Guo, G. Bian, Z. Yuan, H. Lu, T. R. Chang, P. P. Shibayev, M. L. Prokopovych, N. Alidoust, H. Zheng, C. C. Lee, S. M. Huang, R. Sankar, F. Chou, C. H. Hsu, H. T. Jeng, A. Bansil, T. Neupert, V. N. Strocov, H. Lin, S. Jia and M. Zahid Hasan, Sci. Adv. {\bf 1}, e1501092 (2015).

  
  
  
  
  

\bibitem[{\citenamefont{Huang et~al.}(2015{\natexlab{a}})\citenamefont{Huang,
  Zhao, Long, Wang, Chen, Yang, Liang, Xue, Weng, Fang, Dai and Chen}}]{huang2015observation}
\bibinfo{author}{\bibfnamefont{X.}~\bibnamefont{Huang}},
  \bibinfo{author}{\bibfnamefont{L.}~\bibnamefont{Zhao}},
  \bibinfo{author}{\bibfnamefont{Y.}~\bibnamefont{Long}},
  \bibinfo{author}{\bibfnamefont{P.}~\bibnamefont{Wang}},
  \bibinfo{author}{\bibfnamefont{D.}~\bibnamefont{Chen}},
  \bibinfo{author}{\bibfnamefont{Z.}~\bibnamefont{Yang}},
  \bibinfo{author}{\bibfnamefont{H.}~\bibnamefont{Liang}},
  \bibinfo{author}{\bibfnamefont{M.}~\bibnamefont{Xue}},
  \bibinfo{author}{\bibfnamefont{H.}~\bibnamefont{Weng}},
  \bibinfo{author}{\bibfnamefont{Z.}~\bibnamefont{Fang}},
   \bibinfo{author}{\bibfnamefont{X.}~\bibnamefont{Dai}},
   \bibnamefont{and}
   \bibinfo{author}{\bibfnamefont{G.}~\bibnamefont{Chen}},  
  \bibinfo{journal}{Phys. Rev. X} \textbf{\bibinfo{volume}{5}},
  \bibinfo{pages}{031023} (\bibinfo{year}{2015}{\natexlab{a}}).



\bibitem{zhang2016signatures}
C. L. Zhang, S. Y. Xu, I. Belopolski, Z. Yuan, Z. Lin, B. Tong, G. Bian, N. Alidoust, C. C. Lee, S. M. Huang, T. R. Chang, G. Chang, C. H. Hsu, H. T. Jeng, M. Neupane, D. S. Sanchez, H. Zheng, J. Wang, H. Lin, C. Zhang, H. Z. Lu, S. Q. Shen, T. Neupert, M Z. Hasan, and S. Jia, Nat. Commun. {\bf 7},  10735 (2016).



\bibitem{shekhar2015extremely}
C. Shekhar, A. K. Nayak, Y. Sun, M. Schmidt, M. Nicklas, I. Leermakers, U. Zeitler, W. Schnelle, J. Grin, C. Felser, and B. Yan, Nat. Phys. {\bf 8}, 645 (2015).
  
  

\bibitem[{\citenamefont{Nielsen and Ninomiya}(1983)}]{nielsen1983adler}
\bibinfo{author}{\bibfnamefont{H.~B.} \bibnamefont{Nielsen}} \bibnamefont{and}
  \bibinfo{author}{\bibfnamefont{M.}~\bibnamefont{Ninomiya}},
  \bibinfo{journal}{Phys. Lett. B} \textbf{\bibinfo{volume}{130}},
  \bibinfo{pages}{389} (\bibinfo{year}{1983}).

\bibitem[{\citenamefont{Aji}(2012)}]{aji2012adler}
\bibinfo{author}{\bibfnamefont{V.}~\bibnamefont{Aji}},
  \bibinfo{journal}{Phys. Rev. B} \textbf{\bibinfo{volume}{85}},
  \bibinfo{pages}{241101} (\bibinfo{year}{2012}).

\bibitem[{\citenamefont{Son and Spivak}(2013)}]{son2013chiral}
\bibinfo{author}{\bibfnamefont{D. T.}~\bibnamefont{Son}} \bibnamefont{and}
  \bibinfo{author}{\bibfnamefont{B. Z.}~\bibnamefont{Spivak}},
  \bibinfo{journal}{Phys. Rev. B} \textbf{\bibinfo{volume}{88}},
  \bibinfo{pages}{104412} (\bibinfo{year}{2013}).

\bibitem[{\citenamefont{Kim et~al.}(2013)\citenamefont{Kim, Kim, Wang, Sasaki,
  Satoh, Ohnishi, Kitaura, Yang, and Li}}]{kim2013dirac}
\bibinfo{author}{\bibfnamefont{H.-J.} \bibnamefont{Kim}},
  \bibinfo{author}{\bibfnamefont{K.-S.} \bibnamefont{Kim}},
  \bibinfo{author}{\bibfnamefont{J.-F.} \bibnamefont{Wang}},
  \bibinfo{author}{\bibfnamefont{M.}~\bibnamefont{Sasaki}},
  \bibinfo{author}{\bibfnamefont{N.}~\bibnamefont{Satoh}},
  \bibinfo{author}{\bibfnamefont{A.}~\bibnamefont{Ohnishi}},
  \bibinfo{author}{\bibfnamefont{M.}~\bibnamefont{Kitaura}},
  \bibinfo{author}{\bibfnamefont{M.}~\bibnamefont{Yang}}, \bibnamefont{and}
  \bibinfo{author}{\bibfnamefont{L.}~\bibnamefont{Li}},
  \bibinfo{journal}{Phys. Rev. Lett.} \textbf{\bibinfo{volume}{111}},
  \bibinfo{pages}{246603} (\bibinfo{year}{2013}).

\bibitem[{\citenamefont{Hosur and Qi}(2013)}]{hosur2013recent}
\bibinfo{author}{\bibfnamefont{P.}~\bibnamefont{Hosur}} \bibnamefont{and}
  \bibinfo{author}{\bibfnamefont{X.}~\bibnamefont{Qi}},
  \bibinfo{journal}{C R Phys} \textbf{\bibinfo{volume}{14}},
  \bibinfo{pages}{857} (\bibinfo{year}{2013}).
  
  
  

\bibitem{huang2015weyl}
S. M. Huang, S. Y. Xu, I. Belopolski, C. C. Lee, G. Chang, B. Wang, N. Alidoust, G. Bian, M. Neupane, C. Zhang, S. Jia, A. Bansil, H. Lin and M. Z. Hasan, Nat. Commun. {\bf 6}, 7373 (2015).
  
  
  
  

\bibitem[{\citenamefont{Liu et~al.}(2016{\natexlab{a}})\citenamefont{Liu, Yang,
  Sun, Zhang, Peng, Yang, Chen, Zhang, Guo, Prabhakaran, Schmidt, Hussain, Mo, Felser, and Yan}}]{liu2016evolution}
\bibinfo{author}{\bibfnamefont{Z.}~\bibnamefont{Liu}},
  \bibinfo{author}{\bibfnamefont{L.}~\bibnamefont{Yang}},
  \bibinfo{author}{\bibfnamefont{Y.}~\bibnamefont{Sun}},
  \bibinfo{author}{\bibfnamefont{T.}~\bibnamefont{Zhang}},
  \bibinfo{author}{\bibfnamefont{H.}~\bibnamefont{Peng}},
  \bibinfo{author}{\bibfnamefont{H.}~\bibnamefont{Yang}},
  \bibinfo{author}{\bibfnamefont{C.}~\bibnamefont{Chen}},
  \bibinfo{author}{\bibfnamefont{Y.}~\bibnamefont{Zhang}},
  \bibinfo{author}{\bibfnamefont{Y.}~\bibnamefont{Guo}},
  \bibinfo{author}{\bibfnamefont{D.}~\bibnamefont{Prabhakaran}},
  \bibinfo{author}{\bibfnamefont{M.}~\bibnamefont{Schmidt}},
  \bibinfo{author}{\bibfnamefont{Z.}~\bibnamefont{Hussain}},
  \bibinfo{author}{\bibfnamefont{S.-K.}~\bibnamefont{Mo}},
  \bibinfo{author}{\bibfnamefont{C.}~\bibnamefont{Felser}},
  \bibinfo{author}{\bibfnamefont{B.}~\bibnamefont{ Yan}},
  \bibnamefont{and}
  \bibinfo{author}{\bibfnamefont{Y. L.}~\bibnamefont{Chen}},
  \bibinfo{journal}{Nat. Mater.}
  \textbf{\bibinfo{volume}{15}}, \bibinfo{pages}{27}
  (\bibinfo{year}{2016}{\natexlab{a}}).
  
  

  
  \bibitem{Nature Physics 2015}
  S. Y. Xu,	N. Alidoust,	I. Belopolski,	Z. Yuan,	G. Bian,	T. R. Chang, H. Zheng,	V. N. Strocov,	D. S. Sanchez,	G. Chang,	C. Zhang, D. Mou,	Y. Wu,	L. Huang,	C. C. Lee,	S. M. Huang,	B. Wang, A. Bansil,	H. T. Jeng,	T. Neupert,	A. Kaminski,	H. Lin,	S. Jia	and M. Z. Hasan, Nat. Phys. {\bf 11}, 748 (2015).
  
  

\bibitem[{\citenamefont{Zhou et~al.}(2016)\citenamefont{Zhou, Lu, Du, Zhu,
  Zhang, Zhang, Shao, Chen, Wang, Tian, Sun, Wan, Yang, Yang, Zhang and Xing}}]{zhou2016pressure}
\bibinfo{author}{\bibfnamefont{Y.}~\bibnamefont{Zhou}},
  \bibinfo{author}{\bibfnamefont{P.}~\bibnamefont{Lu}},
  \bibinfo{author}{\bibfnamefont{Y.}~\bibnamefont{Du}},
  \bibinfo{author}{\bibfnamefont{X.}~\bibnamefont{Zhu}},
  \bibinfo{author}{\bibfnamefont{G.}~\bibnamefont{Zhang}},
  \bibinfo{author}{\bibfnamefont{R.}~\bibnamefont{Zhang}},
  \bibinfo{author}{\bibfnamefont{D.}~\bibnamefont{Shao}},
  \bibinfo{author}{\bibfnamefont{X.}~\bibnamefont{Chen}},
  \bibinfo{author}{\bibfnamefont{X.}~\bibnamefont{Wang}},
  \bibinfo{author}{\bibfnamefont{M.}~\bibnamefont{Tian}},
  \bibinfo{author}{\bibfnamefont{J.}~\bibnamefont{Sun}},
  \bibinfo{author}{\bibfnamefont{X.}~\bibnamefont{Wan}},
  \bibinfo{author}{\bibfnamefont{Z.}~\bibnamefont{Yang}},
  \bibinfo{author}{\bibfnamefont{W.}~\bibnamefont{Yang}},
  \bibinfo{author}{\bibfnamefont{Y.}~\bibnamefont{Zhang}},
  \bibnamefont{and}
  \bibinfo{author}{\bibfnamefont{D.}~\bibnamefont{Xing}},      
  \bibinfo{journal}{Phys. Rev. Lett.} \textbf{\bibinfo{volume}{117}},
  \bibinfo{pages}{146402} (\bibinfo{year}{2016}).



\bibitem[{\citenamefont{Luo et~al.}(2016)\citenamefont{Luo, Ghimire, Bauer,
  Thompson, and Ronning}}]{luo2016hard}
\bibinfo{author}{\bibfnamefont{Y.}~\bibnamefont{Luo}},
  \bibinfo{author}{\bibfnamefont{N.}~\bibnamefont{Ghimire}},
  \bibinfo{author}{\bibfnamefont{E.}~\bibnamefont{Bauer}},
  \bibinfo{author}{\bibfnamefont{J.}~\bibnamefont{Thompson}}, \bibnamefont{and}
  \bibinfo{author}{\bibfnamefont{F.}~\bibnamefont{Ronning}},
  \bibinfo{journal}{J. Phys. Condens. Matter}
  \textbf{\bibinfo{volume}{28}}, \bibinfo{pages}{055502}
  (\bibinfo{year}{2016}).

\bibitem[{\citenamefont{dos Reis et~al.}(2016)\citenamefont{dos Reis, Wu, Sun,
  Ajeesh, Shekhar, Schmidt, Felser, Yan, and Nicklas}}]{dos2016pressure}
\bibinfo{author}{\bibfnamefont{R. D.}~\bibnamefont{dos Reis}},
  \bibinfo{author}{\bibfnamefont{S. C.}~\bibnamefont{Wu}},
  \bibinfo{author}{\bibfnamefont{Y.}~\bibnamefont{Sun}},
  \bibinfo{author}{\bibfnamefont{M. O.}~\bibnamefont{Ajeesh}},
  \bibinfo{author}{\bibfnamefont{C.}~\bibnamefont{Shekhar}},
  \bibinfo{author}{\bibfnamefont{M.}~\bibnamefont{Schmidt}},
  \bibinfo{author}{\bibfnamefont{C.}~\bibnamefont{Felser}},
  \bibinfo{author}{\bibfnamefont{B.}~\bibnamefont{Yan}}, \bibnamefont{and}
  \bibinfo{author}{\bibfnamefont{M.}~\bibnamefont{Nicklas}},
  \bibinfo{journal}{Phys. Rev. B} \textbf{\bibinfo{volume}{93}},
  \bibinfo{pages}{205102} (\bibinfo{year}{2016}).
  
  
 \bibitem{Martin 1988}
 J. Martin and R. Zum Gruehn, Z. Kristallogr {\bf 182}, 180 (1988).
  
  
  
  
  
  \bibitem{Arnold 2016}
  F. Arnold, C. Shekhar, S. C. Wu, Y. Sun, R. D. d. Reis, N. Kumar, M. Naumann, M. O. Ajeesh, M. Schmidt, A. G. Grushin, J. H. Bardarson, M. Baenitz, D. Sokolov, H. Borrmann, M. Nicklas, C. Felser, E. Hassinger and B. Yan, Nat. Commun. {\bf 7}, 11615 (2016). 
  
  
  
  
  
  
  
  \bibitem{qe}
  P. Giannozzi {\it et~al.}, J. Phys. Condens. Matter {\bf 21}, 395502  (2009).
  
  \bibitem{USPP}
  D. Vanderbilt, Phys. Rev. B {\bf 41}, 7892 (1990).
  
  
  \bibitem{PZ}
  J. P. Perdew and A. Zunger,  Phys. Rev. B {\bf 23},  5048  (1981).
  
  
  \bibitem{Corso}
  A. DalCorso and A. MoscaConte, Phys. Rev. B {\bf 71},  115106  (2005).
  

\bibitem{Boller}
H. Boller and E. Parthé, Acta Crystallogr. {\bf 16}, 1095 (1963).

\bibitem[{\citenamefont{Liu et~al.}(2016{\natexlab{b}})\citenamefont{Liu,
  Richard, Zhao, Chen, and Ding}}]{liu2016comparative}
\bibinfo{author}{\bibfnamefont{H.}~\bibnamefont{Liu}},
  \bibinfo{author}{\bibfnamefont{P.}~\bibnamefont{Richard}},
  \bibinfo{author}{\bibfnamefont{L.}~\bibnamefont{Zhao}},
  \bibinfo{author}{\bibfnamefont{G.}~\bibnamefont{Chen}}, \bibnamefont{and}
  \bibinfo{author}{\bibfnamefont{H.}~\bibnamefont{Ding}},
  \bibinfo{journal}{J. Phys. Condens. Matter}
  \textbf{\bibinfo{volume}{28}}, \bibinfo{pages}{295401}
  (\bibinfo{year}{2016}{\natexlab{b}}).




\bibitem{Petricek}
 V. Petricek,  M. Dusek, and L. Palatinus, Z. Kristallogr. 229(5), 345 (2014).


\bibitem{Xiaoa 2010}
W. Xiaoa, D. Tana, X. Xionga, J. Liub and J. Xuc, Proc. Natl. Acad. Sci. {\bf 107}, 14026 (2010)

\bibitem{Zhao 2015}
J. Zhao, L. Xu, Y. Liu, Z. Yu, C. Li, Y. Wang, and Z. Liu, J. Phys. Chem. C , {\bf 119}, 27657 (2015).


\bibitem{Hong 2016}
F. Hong, B. Yue, N. Hirao, G. Ren, B. Chen and H. K. Mao, Appl. Phys. Lett. {\bf 109}, 241904 (2016).



\bibitem{Lee}
C. C. Lee, S. Y. Xu, S. M. Huang, D. S. Sanchez, I. Belopolski, G. Chang, G. Bian, N. Alidoust, H. Zheng, M. Neupane, B. Wang, A. Bansil, M. Z. Hasan, and H. Lin, Phys. Rev. B {\bf 92}, 235104 (2015). 



\bibitem{anjali}
A. Bera, A. Singh, D. V. S. Muthu, U. V. Waghmare and A. K. Sood, J. Phys. Condens. Matter {\bf 29}, 105403 (2017).


\bibitem{Lifshitz}
I. M. Lifshitz, Sov. Phys. JETP {\bf 11}, 1130 (1960).


\end{thebibliography}

\end{document}